\documentclass[12pt]{article}
\usepackage{epsf}

\newcommand{\be}{\begin{equation}}
\newcommand{\ee}{\end{equation}}
\newcommand{\ppkkd}{\mbox {\boldmath $pp\to K^+\bar K^0 d$}}

\newcommand{\dsmkk}{\frac {d\sigma}{dm_{K\bar K}}}
\newcommand{\dsmkd}{\frac {d\sigma}{dm_{\bar K d}}}
\newcommand{\bfeps}{\mbox {\boldmath $\epsilon$}}
\newcommand{\bfp}{\mbox {\boldmath $p$}}
\newcommand{\bfk}{\mbox {\boldmath $k$}}
\newcommand{\bfq}{\mbox {\boldmath $q$}}
\newcommand{\bfs}{\mbox {\boldmath $S$}}

\newcommand{\bfsigma}{\mbox {\boldmath $\sigma$}}
\newcommand{\dsmo}{d^2\sigma/dm_{K\bar K}d\Omega}
\newcommand{\bflks}{\mbox {\boldmath $L_k=0$}}
\newcommand{\bflkp}{\mbox {\boldmath $L_k=1$}}
\newcommand{\bflqs}{\mbox {\boldmath $L_q=0$}}
\newcommand{\bflqp}{\mbox {\boldmath $L_q=1$}}
\newcommand{\bfkk}{\mbox {\boldmath $K\bar K$}}

\date{}

\begin{document}

\centerline{\large \bf On differential spectra in the reaction \ppkkd }
\centerline{\large \bf in the nearthreshold region}

\vspace{5mm}

\centerline{\bf V.P.Chernyshev, P.V.Fedorets,
                A.E.Kudryavtsev, V.E.Tarasov}

\vspace{3mm}

\centerline{Institute of Theoretical and Experimental Physics}
\centerline{117259 Moscow, Russia}

\vspace{5mm}

{\bf 1}. The reactions with production of the lightest scalar mesons are
presently the subject of experimental study by the ANKE collaboration
at COSY machine at J\"ulich~\cite{1,2}.
Recently the preliminary results on the reaction
\be
pp\to K^+\bar K^0 d
\label{1}
\ee
at the proton-beam energy $T_p =2.645$~GeV, obtained in the current
experiment at ANKE spectrometer~\cite{3,4}, have been reported.
The experiment is going on and has the aim to observe the scalar
$a^+_0$(980) meson ($I^G J^{PC} = 1^- 0^{++}$) in the decay channel
$K\bar K$ and to study its properties. Note that the allowed
$K\bar K$-mass interval in the reaction~(\ref{1}) at a given incident
energy is rather narrow (991.4 $< M_{K\bar K} <$ 1037.3 MeV), i.e.
approximately two times smaller than the expected width of $a_0$ meson
($\Gamma_a\sim$ 80~MeV). Thus, the phase-volume limitations should
already reproduce a resonance-like bump in the $K\bar K$-mass spectrum.
In this situation, the observation of $a_0$ meson in the
reaction~(\ref{1}) appears to be not an easy problem.

The subject of this article is to discuss the possible distributions
for different observables one may expect from the reaction~(\ref{1})
at the incident energy $T_p =2.645$~GeV.
The modern effective meson-nucleon theories of strong interactions
are not able to predict cross sections for production of heavy
mesons ($m_a\sim 1$~GeV) with high accuracy. There exist theoretical
predictions for production rate of $a_0$ meson in the reaction
$NN\to a_0 d$, based on a meson exchange model~\cite{Gri}. There are
also estimations, according to which nonresonance background in
the reaction~(\ref{1}) is expected to be strongly suppressed in
comparison with $a_0$ contribution (see, Ref.~\cite{1}).
However, in the present article we consider the problem in the
model-independent way, making use of the conservation laws for
quantum numbers in the reaction~(\ref{1}).

Below we shall discuss the $K^+\bar K^0$- and $\bar K^0 d$-mass spectra
for this reaction with unpolarized particles at different hypotheses
for the production amplitude. From the experimental point of view the
$K^+\bar K^0$-mass spectrum should be the most sensitive to the possible
contribution of $a^+_0$ resonance in this channel.
On the other hand the $\bar K^0 d$-mass spectrum may be essentially
influenced by strong final-state $\bar K^0 d$ interaction (this question
in details was discussed in Ref.~\cite{Oset}). Some comments on the
angular distributions will be given.

\vspace{3mm}
{\bf 2}. Note that at the incident kinetic energy $T_p =2.645$~GeV the
the reaction~(\ref{1}) is rather close to threshold regime
($Q=\sqrt{s} -m_d -m_{K^+} -m_{\bar K^0}\approx 45.9$~MeV, where
$\sqrt{s}$ is total energy in CMS). Thus, one may expect that
contributions of lower partial waves should dominate. Let us introduce
the following notations:

$\bfp$, $\,L_p$ -- relative 3-momentum and orbital angular momentum,
respectively, of the initial protons;

$\bfk$, $\,L_k$ -- relative 3-momentum and orbital angular momentum,
respectively, of the final $K\bar K$ system with respect to the deutron
in CMS;

$\bfq$, $\,L_q$ -- relative 3-momentum of kaons and orbital angular
momentum, respectively, in the final $K\bar K$ system;

${\bf k}_1$ -- relative momenta in the final $\bar K d$ system;

${\bf k}_2$ -- relative momenta of $K$ meson with respect to the
               final $\bar K d$ system in CMS.

In what follows, the final particles in the CMS of the
reaction~(\ref{1}) are considered to be nonrelativistic and the
corresponding momenta are given by the expressions:
$$
q=\sqrt{\frac{2 m_K m_{\bar K}}{m_K + m_{\bar K}}\,
(m_{K\bar K}\!-\!m_K\!-\!m_{\bar K})}\, ,
\phantom{xx}
k=\sqrt{\frac{2 (m_K + m_{\bar K}) m_d}{m_K +m_{\bar K} +m_d}\,
(\sqrt{s}\!-\!m_d\!-\!m_{K\bar K})}\, ,
$$
\be
k_1=\sqrt{\frac{2 m_d m_{\bar K}}{m_d + m_{\bar K}}\,
(m_{\bar K d}\!-\!m_d\!-\!m_{\bar K})}\, ,
\phantom{xx}
k_2=\sqrt{\frac{2 (m_d + m_{\bar K}) m_K}{m_K +m_{\bar K} +m_d}\,
(\sqrt{s}\!-\!m_K\!-\!m_{\bar K d})}\, .
\label{2}
\ee
Differential cross section may be written as
\be
d\sigma = N\, |M|^2\, k q\, dm_{K\bar K}\, d\Omega_k\, d\Omega_q\,
\phantom{xxxx}
(N=(4\pi)^{-5} p^{-1} s^{-1}).
\label{3}
\ee
Here: $M$ is matrix element; $\Omega_k$ and $\Omega_q$ are solid angles
for the directions of the momenta $\bfk$ and $\bfq$, respectively.

In the simplest approximation, in which the production amplitude $M$
is constant, the mass spectra are given only by phase-space limitations
for the three final particles, i.e.
\be
\dsmkk = N_0\,k\,q \, , \phantom{xxx} \dsmkd = N_0\,k_1\,k_2 \, ,
\label{4}
\ee
where $N_0 = (4\pi)^2 N\,|M|^2 = const\,$. The distributions~(\ref{4})
are shown in Fig.1a and Fig.1b (dotted curves) and are symmetric.

However, the approximation $M=const$ corresponds to the forbidden case
$L_k = L_q = 0$. Note that since the final $K^+\bar K^0 d$ system in the
reaction~(\ref{1}) has isospin 1 the case $\,L_k = L_q = 0\,$
is forbidden due to parity, angular momentum and isospin conservation
laws and Pauli principle. Thus, the possible lowest-partial-wave
contributions correspond to the cases: $\,L_k = 1$, $\,L_q =0\,$ and/or
$\,L_k = 0$, $\,L_q =1\,$. In both these cases the initial $NN$ system
has the total spin $S=1$ and the orbital angular momentum
$L_p=1$ or $L_p=3$ (see also Refs.~\cite{Gri, Kud}).

\vspace{3mm}
{\bf 3. The case \bflkp, \bflqs.
Nonresonance production of \bfkk-system}.
In this case the reaction amplitude should be linear in $\bfk$ and
contain odd powers ($\le 3$) of the initial relative momentum $\bfp$.
This amplitude may be written in the general form
\be
M = a\, (\bfp\cdot\bfs)\, (\bfk\cdot\bfeps^*)
  + b\, (\bfp\cdot\bfk)\, (\bfs\cdot\bfeps^*)
  + c\, (\bfk\cdot\bfs)\, (\bfp\cdot\bfeps^*)
  + d\, (\bfp\cdot\bfs)\, (\bfp\cdot\bfeps^*)\, (\bfk\cdot\bfp)\, .
\label{5}
\ee
Hereafter: $\bfeps$ is the polarization vector of the deuteron and
$\bfs =\phi^T_1\sigma_2\bfsigma\phi_2\,$, where $\phi_1$ and $\phi_2\,$
are the spinors of the initial nucleons.
The coefficients $a$, $b$, $c$ and $d$ are independent complex scalar
amplitudes.\footnote{
The maximal value $L_p=3$ for orbital momentum of the initial nucleons
is taken into account in Eq.~(\ref{4}). All the amplitudes $a$, $b$, $c$
and $d$ may be expressed through vertex constants of some effective
Lagrangian for the reaction~(\ref{1}). This amplitudes may also depend
on the total energy $\sqrt{s}$. One may consider them to be a constants
if $\sqrt{s}$ is fixed.}
In terms of amplitudes $a$, $b$, $c$, i.e. omitting amplitude $d$,
the expression~(\ref{4}) was also discussed in Ref.~\cite{3}.
The matrix element~(\ref{5}), when squared and averaged (summed) over
the polarizations of the initial nucleons (final deuteron), gives
\be
\frac{d^2\sigma}{dm_{K\bar K}\,d\Omega_k}
= 4\pi\, N\,(A + B \cos^2\theta)\, k^3 q\, ,
\label{6}
\ee
where $\theta$ is the CM polar angle of outgoing $K\bar K$ system with
respect to the proton beam in the reaction~(\ref{1}) and
\be
A = \frac{1}{2} \left( |a|^2 +|c|^2\right) p^2 ,
\phantom{xx}
B = \left[ |b|^2 +\frac{1}{2} |b+p^2 d|^2
    + {\rm Re} \left( a^* c + (a+c)^* (b+p^2 d) \right)\right] p^2 .
\label{7}
\ee
The values $A$ and $B$ in Eq.~(\ref{5}) are known if any concrete
model is used. If $a=c=0$ in Eq.~(\ref{5}) then $A=0$ and
$\dsmo_k\sim \cos^2\theta$. On the other hand, $\dsmo_k$ should be flat
with respect to $\cos\theta$ if $B=0$. The latter is valid if $a=b=d=0$
or $c=b=d=0$ in Eq.~(\ref{5}). The amplitude~(\ref{5}) is always
necessarily leads to flat distribution on $\Omega_q$ (angular
distribution of outgoing kaon in the rest frame of the
$K^+\bar K^0$-system. For the $K^+\bar K^0$- and $\bar K^0 d$-mass
distributions one gets
\be
\dsmkk = N_1\,k^3\,q\, , \phantom{xxx}
\dsmkd = N_1 \left[ k^2_1 +
\left(\frac{m_d}{m_{\bar K^0} +m_d}\right)^2\!k^2_2\,\right] k_1 k_2\, ,
\label{8}
\ee
where $N_1=(4\pi)^2\, N\, (A+B/3) =const$. The distributions~(\ref{8})
are shown in Fig.1a and Fig.1b by solid curves 1.

\vspace{3mm}
{\bf 4. The case \bflkp, \bflqs. Resonance production of \bfkk-system.}
Consider now the case of a pure resonance production of $K\bar K$ system
through the chain
\be
pp\to a^+_0 \,d\to K^+\bar K^0\,d\, .
\label{9}
\ee
In this case the expression~(\ref{5}) should be considered as the
amplitude of $a^+_0$-meson production in the reaction $pp\to a^+_0 \,d$.
To obtain the amplitude of the reaction~(\ref{9}) we should multiply
the expression~(\ref{4}) by the $a_0$-meson propagator and by the
constant $g_{aKK}$ of the decay $a^+_0\to K^+\bar K^0$. Note that
angular distributions for resonance production mechanism are identical
to those for nonresonance case, discussed above in Section~3.
For the mass spectra we have
\be
\dsmkk = N_2\, | \Pi_a (m_{K\bar K}) |^2\, k^3 q\, ,
\phantom{xxx}
\dsmkd = \frac{1}{2} N_2\, k_1 k_2\,
\int k^2\, | \Pi_a (m_{K\bar K}) |^2\, dz^{\prime}\, .
\label{10}
\ee
Here: $N_2=(4\pi)^2\, N\, (A+B/3) (g_{aKK}/2m_a)^2 =const$;
$\Pi_a (m) =(m - m_a + i\Gamma(m)/2)^{-1}$ is the nonrelativistic
$a_0$ propagator; $m_a$ is the nominal mass of $a^+_0$ meson;
$z^{\prime}=\theta'$ and $\theta'$ is the angle between the directions
of outgoing $\bar K^0$ and $K^+$ mesons in the rest frame of
$\bar K^0 d$ system.
For the width $\Gamma(m)$ in $\Pi_a(m_{K\bar K})$ we use the analytic
expression (Flatte~\cite{Flatte}):
\be
\Gamma(m) = g_1 q_1 + g\,q\, ,
\label{11}
\ee
where $q_1$ is relative momentum in the $a_0\to \pi +\eta$ decay
channel, taken at $m=m_a$, and $q$ is the relative momentum in the
$K\bar K$ system (see Eqs.~(\ref{2})). We use the parameters
$m_a=998$~MeV, $g_1=0.243$ and $g=0.221$ for $a^+_0$ meson from
Ref~\cite{BNL}.

To calculate the integral in the expression for
$d\sigma/dm_{\bar K d}$~(\ref{10}) one should express the values $k^2$
and $m_{K\bar K}$ in terms of variable $z^{\prime}$. The effective mass
$m_{K\bar K}$ can be expressed through the value $q^2$ according to
Eqs.~(\ref{2}). The expressions for $k^2$ and $q^2$ in terms of the
variable $z'$ are the following:
\be
k^2=k^2_1 +\beta^2\, k^2_2 +\,2\beta\,k_1 k_2\, z'\, ,\phantom{xxx}
q^2=\alpha^2_1\, k^2_1 +\beta^2_1\, k^2_2
-\,2\alpha_1\beta_1\,k_1 k_2\, z'\, ,
\label{12}
\ee
$$
\beta = \frac{m_d}{m_{\bar K} +m_d}\, ,
\phantom{xx}
\alpha_1= \frac{m_K}{m_{\bar K} +m_K}\, ,
\phantom{xx}
\beta_1= \frac{m_{\bar K}\, (m_K + m_{\bar K} +m_d)}
{(m_K + m_{\bar K})\, (m_{\bar K} + m_d)}\, .
$$
The distributions~(\ref{10}) are shown in Fig.1a and Fig.1b
by dashed curves.

\vspace{3mm}
{\bf 5. The case \bflks, \bflqp. Nonresonance production of
$\bf p$-wave \bfkk-system in $\bf s$ wave with respect to deuteron.}
The amplitude of the reaction~(\ref{1}) in this case may be written
in the form~(\ref{5}), where $\bfk$ is substituted by $\bfq$, i.e.
\be
M = a\, (\bfp\cdot\bfs)\, (\bfq\cdot\bfeps^*)
  + b\, (\bfp\cdot\bfq)\, (\bfs\cdot\bfeps^*)
  + c\, (\bfq\cdot\bfs)\, (\bfp\cdot\bfeps^*)
  + d\, (\bfp\cdot\bfs)\, (\bfp\cdot\bfeps^*)\, (\bfq\cdot\bfp)\, .
\label{13}
\ee
The values $a$, $b$, $c$ and $d$ are also taken to be constants.
Using this amplitude, one gets
\be
\frac{d^2\sigma}{dm_{K\bar K}\,d\Omega_q}
= 4\pi\, N\,(A + B \cos^2\theta_1)\, k\,q^3\, ,
\label{14}
\ee
Here: $A$ and $B$ are the constants, given in Eqs.~(\ref{7});
$\,\theta_1\,$ is the angle of outgoing $\bar K^0$ meson with respect
to the proton beam in the rest frame of $\bar K^0 d$ system. If
$a=c=0$ in Eq.~(\ref{13}) then $A=0$ and $\dsmo_q\sim \cos^2\theta_1$.
On the other hand, if $a=b=d=0$ or $c=b=d=0$ then $B=0$ and the
distribution~(\ref{14}) should be flat with respect to $\cos\theta_1$.
In any case the amplitude~(\ref{13}) leads to flat angular
distribution on $\Omega_k$.

The $K^+\bar K^0$- and $\bar K^0 d$-mass distributions are the
following:
\be
\dsmkk = N_1\,k\,q^3\, , \phantom{xxx}
\dsmkd = N_1\,(\,\alpha^2_1\,k^2_1 +\beta^2_1\,k^2_2\,)\, k_1 k_2\, ,
\label{15}
\ee
where $N_1=(4\pi)^2\, N\, (A+B/3) =const$. The values $\alpha_1$ and
$\beta_1$ are given in Eqs.~(\ref{12}).
These distributions are shown in Fig.1a and Fig.1b by solid curves 2.

\vspace{3mm}
{\bf 6}.
Let us here discuss the results. Looking at $K^+\bar K^0$-mass
distributions in Fig.~1a, one can see that this mass spectrum is very
sensitive to the choice of the partial-wave amplitude of the reaction.
The cases $\,L_k = 1$, $\,L_q =0\,$ (solid curve 1 and dashed curve)
and $\,L_k = 0$, $\,L_q =1\,$ (solid curve 2) correspond to strongly
different $K^+\bar K^0$-mass distributions. Generally the experimental
data may correspond to some intermediate case as well as to confirm one
of these two limiting cases. The preliminary results~\cite{4} seem to
confirm the variant with $\,L_k = 1$, $\,L_q =0\,$. This variant looks
like more as the argument in favour of $a^+_0$-resonance hypothesis
than against it. However, since the data~\cite{4} are preliminary and
the experiment is going on, the situation may also change.

Comparing $K^+\bar K^0$-mass spectra for resonance (solid curve 1)
and nonresonance (dashed curve) hypotheses, one can see that these
distributions are not drastically different. That is why to separate
$a^+_0$-resonance and nonresonance mechanisms of the reaction~(\ref{1})
at low incident energy $T_p=2645$~MeV seems to be not an easy problem.

The $\bar K^0 d$-mass spectra in Fig.~1b are less sensitive to the
choice of the production amplitude. The spectra shown by solid curves 1
and 2 are rather different, but in the case of $\,L_k = 1$, $\,L_q =0\,$
the $a^+_0$-resonance and nonresonance hypotheses correspond to
approximately the same distributions (dashed curve and solid curve 1).
The $\bar K^0 d$-mass spectrum is more interesting in connection with
effects of final state interaction (FSI) of $\bar K^0$ mesons with
deutron. This question was thoroughly studied in Ref.~\cite{Oset},
and it was found that the calculated mass spectra were essentially
influenced by strong $\bar K^0 d$ FSI, when the latter was taken into
account. Here the following remark is also possible. Suppose the pure
$a^+_0$-production mechanism in the reaction~(\ref{1}). The FSI
process $a_0 d\to K\bar K d$ leads to the contribution, which looks
like a background for $a_0$. This subprocess may also essentially
contribute to the partial waves with $L_k=0$ and $L_q=1$ in the total
amplitude of the reaction~(\ref{1}) and modify $K\bar K$-mass spectrum.
The observation of the FSI-induced desintegration of $a^+_0$ meson
should be a serious argument in favour of the molecular
hypothesis~\cite{Wein} of $a_0$ meson.

Two of us (A.E.K. and V.E.T.) are thankful to the DFG-RFFI grant for
the financial support.

\newpage
\begin{figure}
\centerline{\epsfxsize 15cm \epsffile{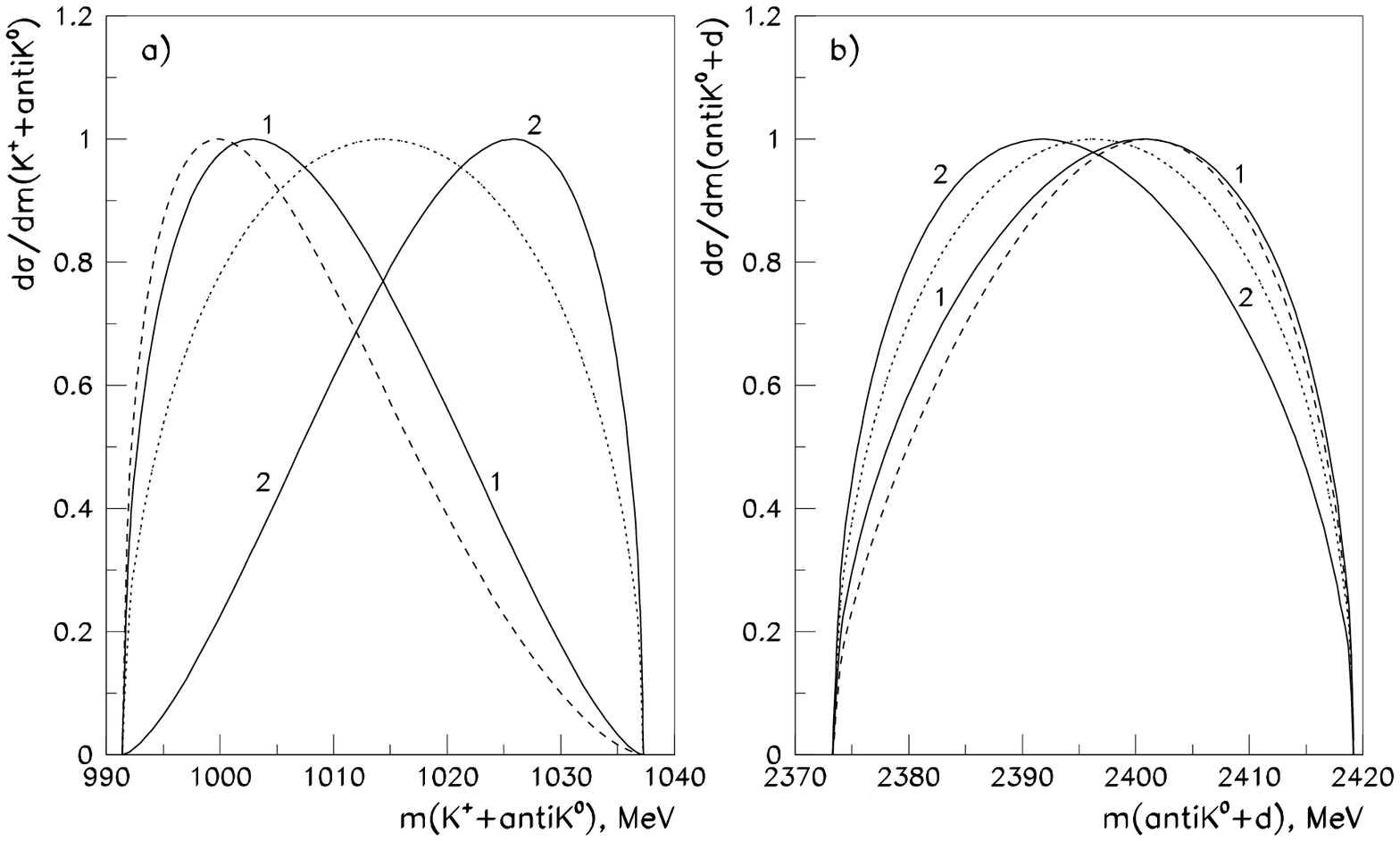}}
\vspace{5mm}
\caption{Mass spectra of $K^+\bar K^0$ (a) and $\bar K^0 d$ (b) systems
       in the reaction~(\ref{1}) at $T_p=2645$~MeV.
       All distributions are normalized to 1 at the maximal values.
       Solid curves 1: nonresonance production with $L_k=1$, $L_q=0$.
       Solid curves 2: nonresonance production with $L_k=0$, $L_q=1$.
       Dashed curves: pure $a^+_0$-meson production ($L_k=1$, $L_q=0$).
       Dotted curves: pure phase-spase distributions
       (this case corresponds to $L_k=L_q=0$ and is forbidden).}
\label{fig1}
\end{figure}

\end{document}